# Multi-party Quantum Private Comparison Protocol Based on Entanglement Swapping of Bell Entangled States

Tian-Yu Ye*

College of Information & Electronic Engineering, Zhejiang Gongshang University, Hangzhou 310018, P.R.China
*E-mail：happyyty@aliyun.com

**Abstract:** Recently, Liu W *et al.* proposed a two-party quantum private comparison (QPC) protocol using entanglement swapping of Bell entangled state (Commun. Theor. Phys. 57(2012)583-588). Subsequently, Liu W J *et al.* pointed out that in Liu W *et al.*'s protocol, the TP can extract the two users' secret inputs without being detected by launching the Bell-basis measurement attack, and suggested the corresponding improvement to mend this loophole (Commun. Theor. Phys. 62(2014)210-214). In this paper, we first point out the information leakage problem toward TP existing in both of the above two protocols, and then suggest the corresponding improvement by using the one-way hash function to encrypt the two users' secret inputs. We further put forward the three-party QPC protocol also based on entanglement swapping of Bell entangled state, and then validate its output correctness and its security in detail. Finally, we generalize the three-party QPC protocol into the multi-party case, which can accomplish arbitrary pair's comparison of equality among *K* users within one execution.

## 1   Introduction

Secure multi-party computation (SMPC), which was first introduced by Yao A C[1] in the millionaire problem, is a basic and important topic in classical cryptography. In Yao A C's millionaire problem, two millionaires wish to know who is richer under the condition of not revealing the genuine amount of asset to each other. Afterward, Boudot F *et al.* [2] constructed an equality comparison protocol to judge whether two millionaires are equally rich. SMPC can be applied into many scenarios such as private bidding and auctions, secret ballot elections, e-commerce, data mining and so on.

As a particular branch of SMPC, classical private comparison (CPC) aims to determine whether two secret inputs from different users are equal or not without disclosing their genuine values. With the development of quantum technology, CPC has been extensively generalized to its quantum counterpart, i.e., quantum private comparison (QPC), whose security is based on the quantum mechanics principles rather than the computation complexity. However, Lo H K[3] pointed out that in a two-party scenario, the equality function cannot be securely evaluated. Under this circumstance, some additional assumptions, for example, a third party (TP), are needed.

The first QPC protocol was proposed by Yang Y G *et al.* [4] using Einstein-Podolsky-Rosen (EPR) pairs with the help of one TP. In the same year, Yang Y G *et al.* [5] proposed the QPC protocol with single photons. The security of these two protocols are essentially based on the one-way hash function. In 2010, Chen X B *et al.* [6] designed the QPC protocol with Greenberger-Horne-Zeilinger (GHZ) states. In 2012, Tseng H Y *et al.* [7] constructed a novel QPC protocol with EPR pairs. In these two protocols, the secret inputs from two users are encrypted with the one-time-pad keys derived from the single-particle measurements. In 2012, Liu W *et al.*[8] proposed the QPC protocol based on entanglement swapping of Bell states (hereafter, this protocol is called as LWC-QPC protocol.) . In this protocol, the secret inputs from two users are encrypted with the one-time-pad keys derived from the Bell-basis measurements after entanglement swapping of the original Bell states. However, Liu W J *et al.*[9] pointed out that in the protocol of Ref.[8], the TP can extract the two users' secret inputs without being detected by launching the Bell-basis measurement attack, and suggested an improved protocol (hereafter, this improved protocol is called as LLCLL-improved-QPC protocol.). Up to now, besides the protocols mentioned above, many other two-party QPC protocols [10-34] have also been designed with different quantum states and quantum technologies.

As to the role of TP, Chen X B *et al.* [6] first introduced the semi-honest model. That is, TP executes the protocol loyally, records all its intermediate computations but might try to reveal the users' secret inputs from the record under the limit that he cannot conspire with the adversary including the dishonest user. However, Yang Y G *et al.* [12] pointed out that this model of semi-honest TP was unreasonable and thought that the reasonable one should be in the following way: TP is allowed to misbehave on his own and also cannot be corrupted by the adversary including the dishonest user. In fact, up to now, this kind of assumption for TP is the most reasonable one.

Suppose that there are *K* users, each of whom has a secret input. They want to know whether their *K* secret inputs are equal or not without disclosing them. If the two-party QPC protocol is adopted to solve this multi-party equality comparison problem, the same two-party QPC protocol has to be executed with $(K-1) \sim K(K-1)/2$ times so that the efficiency is not high enough. In 2013, Chang Y J *et al.* [35]

proposed the first multi-party quantum private comparison (MQPC) protocol with $n$-particle GHZ class states, which can accomplish arbitrary pair's comparison of equality among $K$ users within one execution. Subsequently, the MQPC protocol based on $d$-dimensional basis states and quantum fourier transform [36], and the MQPC protocol based on $n$-level entangled states and quantum fourier transform [37] were constructed. However, there are still few MQPC protocols until now.

In this paper, after carefully investigating the LLCLL-improved-QPC protocol, we find out that it still has an information leakage problem toward TP. Then we suggest an improved strategy for this loophole. We further put forward the three-party QPC protocol also based on entanglement swapping of Bell entangled state and generalize it into the multi-party case accordingly.

## 2 Review of the LLCLL-improved-QPC protocol

For integrity, in this section, a brief review of the LLCLL-improved-QPC protocol is given.

Alice and Bob have two secret integers, $X$ and $Y$, respectively, where $X = \sum_{j=0}^{L-1} x_j 2^j$ and $Y = \sum_{j=0}^{L-1} y_j 2^j$. Here, $x_j, y_j \in \{0,1\}$. They want to know whether $X$ and $Y$ are equal or not with the help of a semi-honest TP.

The LLCLL-improved-QPC protocol can be depicted in the following way:

**Step 1:** Alice/Bob divides her/his binary representation of $X/Y$ into $\lceil L/2 \rceil$ groups $G_1^A, G_2^A, \ldots, G_{\lceil L/2 \rceil}^A$ / $G_1^B, G_2^B, \ldots, G_{\lceil L/2 \rceil}^B$, where each group contains two binary bits. If $L \bmod 2 = 1$, one 0 should be added to $G_{\lceil L/2 \rceil}^A / G_{\lceil L/2 \rceil}^B$ by Alice/Bob.

**Step 2:** Alice/Bob/TP prepares $\lceil L/2 \rceil$ quantum states all in the state of $|\Phi^+\rangle_{A_1 A_2} / |\Phi^+\rangle_{B_1 B_2} / |\Phi^+\rangle_{T_1 T_2}$. Afterward, Alice/Bob/TP picks out the first particle from each state to form an ordered sequence $S_1^A / S_1^B / S_1^T$. The remaining second particle from each state automatically forms the other ordered sequence $S_2^A / S_2^B / S_2^T$.

**Step 3\*:** Alice/TP prepares $L'$ decoy photons randomly in one of the four states $\{|0\rangle, |1\rangle, |+\rangle, |-\rangle\}$ to form sequence $D_A / D_T$. Then, Alice/TP randomly inserts $D_A / D_T$ into $S_2^A / S_2^T$ to obtain $S_2^{A^*} / S_2^{T^*}$. Afterward, Alice and TP exchange $S_2^{A^*}$ and $S_2^{T^*}$ between them. To check the security of the TP-Alice channel, Alice and TP implement the following procedures after Alice receives $S_2^{T^*}$: (1) TP tells Alice the positions and the measurement bases of decoy photons in $S_2^{T^*}$; (2) Alice uses the measurement bases TP told to measure the decoy photons in $S_2^{T^*}$ and informs TP of her measurement results; (3) TP computes the error rate by comparing the initial states of the decoy photons in $S_2^{T^*}$ with Alice's measurement results. If the error rate is low enough, they will continue the next step and Alice will drop out the decoy photons in $S_2^{T^*}$; otherwise, they will halt the communication.

**Step 4:** For $j = 1, 2, \ldots, \lceil L/2 \rceil$, Alice performs the Bell-basis measurement on each pair in $(S_1^A, S_2^T)$ and obtains the corresponding measurement result $M_j^A$. If $M_j^A$ is $|\Phi^+\rangle/|\Phi^-\rangle/|\Psi^+\rangle/|\Psi^-\rangle$, then $R_j^A = 00/01/10/11$. Consequently, the corresponding pair in $(S_1^T, S_2^A)$ in TP's hands is collapsed into one of the four Bell states. These $\lceil L/2 \rceil$ collapsed Bell states in TP's hands are denoted by $(S_1^{T'}, S_2^{T'})$.

**Step 5\*:** Bob prepares $L'$ decoy photons randomly in one of the four states $\{|0\rangle, |1\rangle, |+\rangle, |-\rangle\}$ to form sequence $D_B$ and randomly inserts $D_B$ into $S_2^B$ to obtain $S_2^{B^*}$. Then, Bob and TP exchange $S_2^{B^*}$ and $S_2^{A^*}$ between them. After TP receives $S_2^{B^*}$, TP and Bob check the security of the Bob-TP channel with the same method as that in Step 3\*. On the other hand, after Bob receives $S_2^{A^*}$, Bob makes one-time eavesdropping check for the Alice-TP channel and the TP-Bob channel with Alice by checking the decoy photons in $S_2^{A^*}$. If all quantum channels are secure, Bob and TP will discard the decoy photons and continue the next step.

**Step 6:** For $j = 1, 2, \ldots, \lceil L/2 \rceil$, Bob performs the Bell-basis measurement on each pair in $(S_1^B, S_2^{T'})$ and obtains the corresponding measurement result $M_j^B$. If $M_j^B$ is $|\Phi^+\rangle/|\Phi^-\rangle/|\Psi^+\rangle/|\Psi^-\rangle$, then $R_j^B = 00/01/10/11$. Consequently, the corresponding pair in $(S_1^{T'}, S_2^B)$ in TP's hands is collapsed into one of the four Bell states. TP also performs the Bell-basis measurement on each pair in $\left(S_1^{T_-}, S_2^B\right)$ and obtains the corresponding measurement result $M_j^T$. If $M_j^T$ is $\left|\Phi^+\right\rangle / \left|\Phi^-\right\rangle / \left|\Psi^+\right\rangle / \left|\Psi^-\right\rangle$, then $R_j^T = \left(r_j^{T_1} r_j^{T_2}\right) = 00/01/10/11$.

**Step 7:** For $j = 1, 2, \ldots, \lceil L/2 \rceil$, Alice and Bob calculate $R_j = \left(R_j^A \oplus G_j^A\right) \oplus \left(R_j^B \oplus G_j^B\right) = \left(r_j^1 r_j^2\right)$, and send $R_j$ to TP. Then, TP calculates $R = \sum_{j=1}^{\lceil L/2 \rceil} \left(\left(r_j^1 \oplus r_j^{T_1}\right) + \left(r_j^2 \oplus r_j^{T_2}\right)\right)$. Without loss of generality, we assume that Alice needs to send the result of $R_j^A \oplus G_j^A$ to Bob for calculating $R_j$.

**Step 8:** TP sends $R$ to Alice and Bob. If $R = 0$, Alice and Bob conclude that $X = Y$; otherwise, they know that $X \neq Y$.

Note that the LLCLL-improved-QPC protocol only makes change for Steps 3 and 5 of LWC-QPC protocol. Steps 1,2,4,6,7 and 8 of LWC-QPC protocol are kept unchanged.

## 3 The information leakage problem and the corresponding improvement

In this Section, we first point out the information leakage problem in the LLCLL-improved-QPC protocol in Sec.3.1, then suggest the corresponding improvement in Sec.3.2.

### 3.1 The information leakage problem

The protocol involves many different parameters, including Alice's two-bit input $G_j^A$, Bob's two-bit input $G_j^B$, Alice's measurement result $M_j^A$, the coding of Alice's measurement result $R_j^A$, Bob's measurement result $M_j^B$, the coding of Bob's measurement result $R_j^B$, TP's measurement result $M_j^T$, the coding of TP's measurement result $R_j^T$, the result of $\left(R_j^A \oplus G_j^A\right) \oplus \left(R_j^B \oplus G_j^B\right)$ (i.e., $R_j$) and the result of $\left(r_j^1 \oplus r_j^{T_1}\right) + \left(r_j^2 \oplus r_j^{T_2}\right)$ (i.e., $R_j'$). The relations among these different parameters when $G_j^A = 00$ are listed in Table 1 (See appendix). It is easy to find out that $R_j'$ totally has three different kinds of value, i.e., 0, 1 and 2. When $R_j' = 0$, we have $G_j^A = G_j^B$; otherwise, it follows $G_j^A \neq G_j^B$. After deducing all the relations among these different parameters when $G_j^A = 01$, $G_j^A = 10$ and $G_j^A = 11$, respectively, we can further summarize the relations between $R_j'$ and $G_j^A$, $G_j^B$, which are shown in Table 2 (See appendix). From Table 2, it is easy to know that when $R_j' = 0$, $\left(G_j^A, G_j^B\right)$ may be (00,00), (01,01), (10,10) or (11,11); when $R_j' = 1$, $\left(G_j^A, G_j^B\right)$ may be (00,01), (01,00), (10,11), (11,10), (00,10), (01,11), (10,00) or (11,01); and when $R_j' = 2$, $\left(G_j^A, G_j^B\right)$ may be (00,11), (01,10), (10,01) or (11,00). Furthermore, when $G_j^A = G_j^B$, there are totally four kinds of $\left(G_j^A, G_j^B\right)$; and when $G_j^A \neq G_j^B$, there are totally twelve kinds of $\left(G_j^A, G_j^B\right)$. As a result, when $R_j' = 1$, the eight possible kinds of $\left(G_j^A, G_j^B\right)$ include 3 bits for TP, which means that $\log_2 3 - 1$ bit information has been leaked out to TP; and when $R_j' = 2$, the four possible kinds of $\left(G_j^A, G_j^B\right)$ include 2 bits for TP, which means that $\log_2 3$ bit information has been leaked out to TP. This protocol has an information leakage problem toward TP indeed.

### 3.2 The corresponding improvement

In order to avoid the information leakage problem toward TP, we should make TP get nothing about $G_j^A$ and $G_j^B$ when $R_j' \neq 0$. In this Subsection, we give an improvement to mend this loophole. In order to retain the main features of the LWC-QPC protocol, we make as few modifications as possible. The LWC-QPC protocol should be modified as follows:

**Step 1#:** Similar to the QPC protocols of Refs.[4-5], Alice and Bob share a secret one-way hash function $H$ in advance. Here, the one-way hash function is defined as: $H: \{0,1\}^L \to \{0,1\}^N$, where $L$ is the length of the secret inputs and $N$ is the length of the hash values of the secret inputs. The hash values of $X$ and $Y$ are $H(X) = X^\# = \left(x_{N-1}^\#, x_{N-2}^\#, \ldots, x_0^\#\right)$ and $H(Y) = Y^\# = \left(y_{N-1}^\#, y_{N-2}^\#, \ldots, y_0^\#\right)$, respectively. Alice/Bob divides her/his binary representation of $X^\# / Y^\#$ into $\lceil N/2 \rceil$ group $G_1^{A^\#}, G_2^{A^\#}, \ldots, G_{\lceil N/2 \rceil}^{A^\#}$ / $G_1^{B^\#}, G_2^{B^\#}, \ldots, G_{\lceil N/2 \rceil}^{B^\#}$, where each group contains two binary bits. If $N \bmod 2 = 1$, one 0 should be added to $G_{\lceil N/2 \rceil}^{A^\#} / G_{\lceil N/2 \rceil}^{B^\#}$ by Alice/Bob.

**Step 2#:** Alice/Bob/TP prepares $\lceil N/2 \rceil$ quantum states all in the state of $\left|\Phi^+\right\rangle_{A_1 A_2} / \left|\Phi^+\right\rangle_{B_1 B_2} / \left|\Phi^+\right\rangle_{T_1 T_2}$.

Afterward, Alice, Bob and TP do the same thing as that in Step 2 of the LWC-QPC protocol.

**Step $3^\#$, $4^\#$, $5^\#$ and $6^\#$:** These Steps here are the same as those of the LWC-QPC protocol.

**Step $7^\#$:** For $j = 1, 2, \ldots, \lceil N/2 \rceil$, Alice and Bob calculate $R_j = \left( R_j^A \oplus G_j^{A^\#} \right) \oplus \left( R_j^B \oplus G_j^{B^\#} \right) = \left( r_j^1 r_j^2 \right)$, and send $R_j$ to TP. Then, TP calculates $R = \sum_{j=1}^{\lceil N/2 \rceil} \left( \left( r_j^1 \oplus r_j^{T_1} \right) + \left( r_j^2 \oplus r_j^{T_2} \right) \right)$. Without loss of generality, we assume that Alice needs to send the result of $R_j^A \oplus G_j^{A^\#}$ to Bob for calculating $R_j$.

**Step $8^\#$:** This Step here is the same as that of the LWC-QPC protocol.

Compared with the LWC-QPC protocol, in the above improvement, we add the encryption process for Alice and Bob' secret inputs with a one-way hash function to enhance their privacy. Similar to the LWC-QPC protocol, in the above improvement, TP can also obtain the relations between $R_j'$ and $G_j^{A^\#}$, $G_j^{B^\#}$, which are shown in Table 3 (See appendix). However, the one-way property of the hash function can guarantee that knowing $G_j^{A^\#}$ and $G_j^{B^\#}$ is still helpless to deduce $G_j^A$ and $G_j^B$. As a result, TP cannot get the relations between $R_j'$ and $G_j^A$, $G_j^B$ when $R_j' \neq 0$. Therefore, none of information about Alice and Bob' secret inputs have been leaked out to TP when $R_j' \neq 0$. It can be concluded that using a one-way hash function to encrypt Alice and Bob' secret inputs beforehand helps overcome the information leakage problem toward TP.

It should be further emphasized that in order to retain the main features of the LWC-QPC protocol as many as possible, the above improvement still adopts the same eavesdropping check methods to those used in the LWC-QPC protocol. Because the encryption process for Alice and Bob' secret inputs with a one-way hash function can automatically resist the Bell-basis measurement attack from TP suggested by Liu W J et al. [9], it is not necessary for the above improvement to employ the decoy photon eavesdropping check methods any more.

## 4 The three-party QPC protocol based on entanglement swapping of Bell entangled states

In this Section, by utilizing the above analysis, we suggest the three-party QPC protocol based on entanglement swapping of Bell entangled states in Sec.4.1 first, then analyze its correctness and security in Sec.4.2.

### 4.1 The three-party QPC protocol

Alice, Bob and Charlie have three secret integers, $X$, $Y$ and $Z$, respectively, where $X = \sum_{j=0}^{L-1} x_j 2^j$, $Y = \sum_{j=0}^{L-1} y_j 2^j$ and $Z = \sum_{j=0}^{L-1} z_j 2^j$. Here, $x_j, y_j, z_j \in \{0,1\}$. They want to know whether every two of $X$, $Y$ and $Z$ are equal or not with the help of a semi-honest TP. They achieve the equality comparison of every two secret integers by implementing the following steps.

**Step 1: Preparation.**

(1) Similar to the QPC protocols of Refs.[4-5], Alice, Bob and Charlie share a secret one-way hash function $H$ in advance. The hash values of $X$, $Y$ and $Z$ are $H(X) = X^\# = \left( x_{N-1}^\#, x_{N-2}^\#, \ldots, x_0^\# \right)$, $H(Y) = Y^\# = \left( y_{N-1}^\#, y_{N-2}^\#, \ldots, y_0^\# \right)$ and $H(Z) = Z^\# = \left( z_{N-1}^\#, z_{N-2}^\#, \ldots, z_0^\# \right)$, respectively. Alice/Bob/Charlie divides her/his/her binary representation of $X^\# / Y^\# / Z^\#$ into $\lceil N/2 \rceil$ groups $G_1^{A^\#}, G_2^{A^\#}, \ldots, G_{\lceil N/2 \rceil}^{A^\#} / G_1^{B^\#}, G_2^{B^\#}, \ldots, G_{\lceil N/2 \rceil}^{B^\#} / G_1^{C^\#}, G_2^{C^\#}, \ldots, G_{\lceil N/2 \rceil}^{C^\#}$, where each group contains two binary bits. If $N \bmod 2 = 1$, one 0 should be added to $G_{\lceil N/2 \rceil}^{A^\#} / G_{\lceil N/2 \rceil}^{B^\#} / G_{\lceil N/2 \rceil}^{C^\#}$ by Alice/Bob/ Charlie.

(2) Alice/Bob/Charlie/TP prepares $\lceil N/2 \rceil$ quantum states all in the state of $\left| \Phi^+ \right\rangle_{A_1 A_2} / \left| \Phi^+ \right\rangle_{B_1 B_2} / \left| \Phi^+ \right\rangle_{C_1 C_2} / \left| \Phi^+ \right\rangle_{T_1 T_2}$. Afterward, Alice/Bob/Charlie/TP picks out the first particle from each state to form an ordered sequence $S_1^A / S_1^B / S_1^C / S_1^T$. The remaining second particle from each state automatically forms the other ordered sequence $S_2^A / S_2^B / S_2^C / S_2^T$.

(3) For the security check, Alice/TP prepares a sequence of $L'$ quantum states all in the state of $\left| \Phi^+ \right\rangle$ again, which is denoted as $D_{A'} / D_{T'}$. Then Alice/TP inserts the first and the second particles of each Bell state in $D_{A'} / D_{T'}$ into $S_1^A / S_1^T$ and $S_2^A / S_2^T$ at the same positions, respectively. Accordingly, Alice/TP obtains $S_1^{A'} / S_1^{T'}$ and $S_2^{A'} / S_2^{T'}$. Then, Alice and TP exchange $S_2^{A'}$ and $S_2^{T'}$ between them. To ensure the

transmission security of Alice-TP/TP-Alice quantum channel, the entanglement correlation between two different particles of each Bell state in $D_A / D_T$ is used to check whether there is an eavesdropper or not. If there is no eavesdropper, Alice and TP drop out the sample particles, and implement the next step.

(4) For $j = 1, 2, \ldots, \lceil N/2 \rceil$, Alice performs the Bell-basis measurement on each pair in $\left(S_1^A, S_2^T\right)$ and obtains the corresponding measurement result $M_j^A$. If $M_j^A$ is $|\Phi^+\rangle / |\Phi^-\rangle / |\Psi^+\rangle / |\Psi^-\rangle$, then $R_j^A = 00/01/10/11$. Consequently, the corresponding pair in $\left(S_1^T, S_2^A\right)$ in TP's hands is collapsed into one of the four Bell states. These $\lceil N/2 \rceil$ collapsed Bell states in TP's hands are denoted by $\left(S_1^{T^1}, S_2^{T^1}\right)$.

**Step 2 : The first round comparison.**

(1) Bob/TP prepares a sequence of $L$ quantum states all in the state of $|\Phi^+\rangle$ to guarantee the security for the exchange of $S_2^B$ and $S_2^{T^1}$. If there is no eavesdropper, Bob and TP drop out the sample particles, and implement the next step.

(2) For $j = 1, 2, \ldots, \lceil N/2 \rceil$, Bob performs the Bell-basis measurement on each pair in $\left(S_1^B, S_2^{T^1}\right)$ and obtains the corresponding measurement result $M_j^B$. If $M_j^B$ is $|\Phi^+\rangle / |\Phi^-\rangle / |\Psi^+\rangle / |\Psi^-\rangle$, then $R_j^B = 00/01/10/11$. Consequently, the corresponding pair in $\left(S_1^{T^1}, S_2^B\right)$ in TP's hands is collapsed into one of the four Bell states. TP also performs the Bell-basis measurement on each pair in $\left(S_1^{T^1}, S_2^B\right)$ and obtains the corresponding measurement result $M_j^{T^1}$. If $M_j^{T^1}$ is $|\Phi^+\rangle / |\Phi^-\rangle / |\Psi^+\rangle / |\Psi^-\rangle$, then $R_j^{T^1} = \left(r_j^{T_1^1} r_j^{T_2^1}\right) = 00/01/10/11$. These $\lceil N/2 \rceil$ collapsed Bell states in TP's hands are denoted by $\left(S_1^{T^2}, S_2^{T^2}\right)$.

(3) For $j = 1, 2, \ldots, \lceil N/2 \rceil$, Alice and Bob cooperate to calculate $R_j^{AB} = \left(R_j^A \oplus G_j^{A^\#}\right) \oplus \left(R_j^B \oplus G_j^{B^\#}\right) = \left(r_j^{AB_1} r_j^{AB_2}\right)$, and send $R_j^{AB}$ to TP. Without loss of generality, we assume that Alice needs to send the result of $R_j^A \oplus G_j^{A^\#}$ to Bob for calculating $R_j^{AB}$. Then, TP calculates $R_j^{AB'} = \left(r_j^{AB_1} \oplus r_j^{T_1^1}\right) + \left(r_j^{AB_2} \oplus r_j^{T_2^1}\right)$ and $R^{AB} = \sum_{j=1}^{\lceil N/2 \rceil} R_j^{AB'}$. Afterward, TP publishes $R^{AB}$ to Alice and Bob. If $R^{AB} = 0$, Alice and Bob conclude that $X = Y$; otherwise, they know that $X \neq Y$.

**Step 3 : The second round comparison.**

(1) Charlie/TP prepares a sequence of $L$ quantum states all in the state of $|\Phi^+\rangle$ to guarantee the security for the exchange of $S_2^C$ and $S_2^{T^2}$. If there is no eavesdropper, Charlie and TP drop out the sample particles, and implement the next step.

(2) For $j = 1, 2, \ldots, \lceil N/2 \rceil$, Charlie performs the Bell-basis measurement on each pair in $\left(S_1^C, S_2^{T^2}\right)$ and obtains the corresponding measurement result $M_j^C$. If $M_j^C$ is $|\Phi^+\rangle / |\Phi^-\rangle / |\Psi^+\rangle / |\Psi^-\rangle$, then $R_j^C = 00/01/10/11$. Consequently, the corresponding pair in $\left(S_1^{T^2}, S_2^C\right)$ in TP's hands is collapsed into one of the four Bell states. TP also performs the Bell-basis measurement on each pair in $\left(S_1^{T^2}, S_2^C\right)$ and obtains the corresponding measurement result $M_j^{T^2}$. If $M_j^{T^2}$ is $|\Phi^+\rangle / |\Phi^-\rangle / |\Psi^+\rangle / |\Psi^-\rangle$, then $R_j^{T^2} = \left(r_j^{T_1^2} r_j^{T_2^2}\right) = 00/01/10/11$.

(3) For $j = 1, 2, \ldots, \lceil N/2 \rceil$, Alice, Bob and Charlie cooperate to calculate $R_j^{BC} = R_j^A \oplus \left(R_j^B \oplus G_j^{B^\#}\right) \oplus \left(R_j^C \oplus G_j^{C^\#}\right) = \left(r_j^{BC_1} r_j^{BC_2}\right)$, and send $R_j^{BC}$ to TP. Without loss of generality, assume that Alice and Bob send $R_j^A$ and the result of $R_j^B \oplus G_j^{B^\#}$ to Charlie for calculating $R_j^{BC}$, respectively. Then, TP calculates $R_j^{BC'} = \left(r_j^{BC_1} \oplus r_j^{T_1^2}\right) + \left(r_j^{BC_2} \oplus r_j^{T_2^2}\right)$ and $R^{BC} = \sum_{j=1}^{\lceil N/2 \rceil} R_j^{BC'}$.

In the meanwhile, for $j = 1, 2, \ldots, \lceil N/2 \rceil$, Alice, Bob and Charlie cooperate to calculate $R_j^{AC} = \left(R_j^A \oplus G_j^{A^\#}\right) \oplus R_j^B \oplus \left(R_j^C \oplus G_j^{C^\#}\right) = \left(r_j^{AC_1} r_j^{AC_2}\right)$, and send $R_j^{AC}$ to TP. Without loss of generality, assume that Alice and Bob send the result of $R_j^A \oplus G_j^{A^\#}$ and $R_j^B$ to Charlie for calculating $R_j^{AC}$, respectively. Then, TP calculates $R_j^{AC'} = \left(r_j^{AC_1} \oplus r_j^{T_1^2}\right) + \left(r_j^{AC_2} \oplus r_j^{T_2^2}\right)$ and $R^{AC} = \sum_{j=1}^{\lceil N/2 \rceil} R_j^{AC'}$.

Finally, TP sends $R^{BC}$ to Bob and Charlie. If $R^{BC} = 0$, Bob and Charlie conclude that $Y = Z$; otherwise, they know that $Y \neq Z$. On the other hand, TP sends $R^{AC}$ to Alice and Charlie. If $R^{AC} = 0$, Alice and Charlie conclude that $X = Z$; otherwise, they know that $X \neq Z$. Until now, the protocol is finished.

For clarity, the entanglement swapping process of Bell states among the four participants of the above three-party QPC protocol is further shown in Fig.1 (See appendix).

### 4.2 Analysis

We analyze the above three-party QPC protocol from the aspects of correctness and security here.

#### 4.2.1 Correctness

There are three cases of correctness need to be discussed in total.

**Case 1: The quality comparison of Alice and Bob's secret inputs**

As for the quality comparison of $X$ and $Y$, Alice and Bob need to calculate $R_j^{AB} = \left(R_j^A \oplus G_j^{A^\#}\right) \oplus \left(R_j^B \oplus G_j^{B^\#}\right) = \left(r_j^{AB_1} r_j^{AB_2}\right)$. Moreover, TP needs to calculate $R_j^{AB'} = \left(r_j^{AB_1} \oplus r_j^{T_1^1}\right) + \left(r_j^{AB_2} \oplus r_j^{T_2^1}\right)$ and $R^{AB} = \sum_{j=1}^{\lceil N/2 \rceil} R_j^{AB'}$. According to Fig.1, the following evolution is satisfied:

$$\begin{cases} R_{A_1A_2} \oplus R_{T_1T_2} = R_{A_1T_2} \oplus R_{T_1A_2} \\ R_{T_1A_2} \oplus R_{B_1B_2} = R_{T_1B_2} \oplus R_{B_1A_2} \end{cases} \Rightarrow R_{A_1A_2} \oplus R_{T_1T_2} = R_{A_1T_2} \oplus \left(R_{B_1B_2} \oplus R_{T_1B_2} \oplus R_{B_1A_2}\right)$$

$$\Rightarrow 00 = R_j^A \oplus R_j^B \oplus R_j^{T^1}$$

$$\Rightarrow G_j^{A^\#} \oplus G_j^{B^\#} = \left(R_j^A \oplus G_j^{A^\#}\right) \oplus \left(R_j^B \oplus G_j^{B^\#}\right) \oplus R_j^{T^1} = R_j^{AB} \oplus R_j^{T^1}$$

$$\Rightarrow R_j^{AB'} = \begin{cases} 0, & \text{if } G_j^{A^\#} = G_j^{B^\#}; \\ 1 \text{ or } 2, & \text{if } G_j^{A^\#} \neq G_j^{B^\#}. \end{cases}$$

$$\Rightarrow R^{AB} = \sum_{j=1}^{\lceil N/2 \rceil} R_j^{AB'} = \begin{cases} 0, & \text{if } X = Y; \\ others, & \text{if } X \neq Y. \end{cases} \quad (1)$$

Therefore, the quality comparison result of $X$ and $Y$ in the above three-party QPC protocol is correct.

**Case 2: The quality comparison of Bob and Charlie's secret inputs**

As for the quality comparison of $Y$ and $Z$, Alice, Bob and Charlie need to calculate $R_j^{BC} = R_j^A \oplus \left(R_j^B \oplus G_j^{B^\#}\right) \oplus \left(R_j^C \oplus G_j^{C^\#}\right) = \left(r_j^{BC_1} r_j^{BC_2}\right)$. Moreover, TP needs to calculate $R_j^{BC'} = \left(r_j^{BC_1} \oplus r_j^{T_1^2}\right) + \left(r_j^{BC_2} \oplus r_j^{T_2^2}\right)$ and $R^{BC} = \sum_{j=1}^{\lceil N/2 \rceil} R_j^{BC'}$. According to Fig.1, the following evolution is satisfied:

$$\begin{cases} R_{A_1A_2} \oplus R_{T_1T_2} = R_{A_1T_2} \oplus R_{T_1A_2} \\ R_{T_1A_2} \oplus R_{B_1B_2} = R_{T_1B_2} \oplus R_{B_1A_2} \\ R_{T_1B_2} \oplus R_{C_1C_2} = R_{T_1C_2} \oplus R_{C_1B_2} \end{cases} \Rightarrow R_{A_1A_2} \oplus R_{T_1T_2} = R_{A_1T_2} \oplus \left(R_{B_1B_2} \oplus \left(R_{B_1A_2} \oplus \left(R_{C_1C_2} \oplus R_{T_1C_2} \oplus R_{C_1B_2}\right)\right)\right)$$

$$\Rightarrow 00 = R_j^A \oplus R_j^B \oplus R_j^C \oplus R_j^{T^2}$$

$$\Rightarrow G_j^{B^\#} \oplus G_j^{C^\#} = R_j^A \oplus \left(R_j^B \oplus G_j^{B^\#}\right) \oplus \left(R_j^C \oplus G_j^{C^\#}\right) \oplus R_j^{T^2} = R_j^{BC} \oplus R_j^{T^2}$$

$$\Rightarrow R_j^{BC'} = \begin{cases} 0, & \text{if } G_j^{B^\#} = G_j^{C^\#}; \\ 1 \text{ or } 2, & \text{if } G_j^{B^\#} \neq G_j^{C^\#}. \end{cases}$$

$$\Rightarrow R^{BC} = \sum_{j=1}^{\lceil N/2 \rceil} R_j^{BC'} = \begin{cases} 0, & \text{if } Y = Z; \\ others, & \text{if } Y \neq Z. \end{cases} \quad (2)$$

Therefore, the quality comparison result of $Y$ and $Z$ in the above three-party QPC protocol is correct.

**Case 3: The quality comparison of Alice and Charlie's secret inputs**

As for the quality comparison of $X$ and $Z$, Alice, Bob and Charlie need to calculate $R_j^{AC} = \left(R_j^A \oplus G_j^{A^\#}\right) \oplus R_j^B \oplus \left(R_j^C \oplus G_j^{C^\#}\right) = \left(r_j^{AC_1} r_j^{AC_2}\right)$. Moreover, TP needs to calculate

$R_j^{AC'} = \left(r_j^{AC_1} \oplus r_j^{T_1^2}\right) + \left(r_j^{AC_2} \oplus r_j^{T_2^2}\right)$ and $R^{AC} = \sum_{j=1}^{\lceil N/2 \rceil} R_j^{AC'}$. According to Fig.1, the following evolution is satisfied:

$$\begin{cases} R_{A_1A_2} \oplus R_{T_1T_2} = R_{A_1T_2} \oplus R_{T_1A_2} \\ R_{T_1A_2} \oplus R_{B_1B_2} = R_{T_1B_2} \oplus R_{B_1A_2} \\ R_{T_1B_2} \oplus R_{C_1C_2} = R_{T_1C_2} \oplus R_{C_1B_2} \end{cases} \Rightarrow R_{A_1A_2} \oplus R_{T_1T_2} = R_{A_1T_2} \oplus \left(R_{B_1B_2} \oplus \left(R_{B_1A_2} \oplus \left(R_{C_1C_2} \oplus R_{T_1C_2} \oplus R_{C_1B_2}\right)\right)\right)$$

$$\Rightarrow 00 = R_j^A \oplus R_j^B \oplus R_j^C \oplus R_j^{T^2}$$

$$\Rightarrow G_j^{A^\#} \oplus G_j^{C^\#} = \left(R_j^A \oplus G_j^{A^\#}\right) \oplus R_j^B \oplus \left(R_j^C \oplus G_j^{C^\#}\right) \oplus R_j^{T^2} = R_j^{AC} \oplus R_j^{T^2}$$

$$\Rightarrow R_j^{AC'} = \begin{cases} 0, & \text{if } G_j^{A^\#} = G_j^{C^\#}; \\ 1 \text{ or } 2, & \text{if } G_j^{A^\#} \neq G_j^{C^\#}. \end{cases}$$

$$\Rightarrow R^{AC} = \sum_{j=1}^{\lceil N/2 \rceil} R_j^{AC'} = \begin{cases} 0, & \text{if } X = Z; \\ \text{others}, & \text{if } X \neq Z. \end{cases} \quad (3)$$

Therefore, the quality comparison result of $X$ and $Z$ in the above three-party QPC protocol is correct.

**4.2.2 Security**

As far as the security is concerned, all of the outside attack, the participant attack and the information leakage problem should be taken into account.

**Case 1: Outside attack**

We analyze the possibility for an outside eavesdropper to get information about $X$, $Y$ and $Z$.

In Step 1(3)/2(1)/3(1), TP and Alice/ Bob/ Charlie exchange two quantum state sequences in their respective hands. However, same to the LWC-QPC protocol, the entanglement correlation between two different particles of each Bell state is used to detect the eavesdropping behavior from an outside attacker. It has been widely accepted that several famous attacks, such as the intercept-resend attack, the measure-resend attack and the entangle-measure attack *et al.*, are invalid to this eavesdropping check method [38-41]. Moreover, except Steps 1(3), 2(1) and 3(1), there is no chance for an eavesdropper to steal as no transmission for quantum states occurs.

In addition, in Steps 2(3) and 3(3), there are classical information transmissions. Suppose that the outside attacker is powerful enough to get all the transmitted classical information. In Step 2(3), the outside attacker obtains the result of $R_j^A \oplus G_j^{A^\#}$ when Alice sends it out to Bob and the result of $R_j^B \oplus G_j^{B^\#}$ when Bob sends $R_j^{AB}$ out to TP. However, as she has no knowledge about the one-time-pad keys $R_j^A$ and $R_j^B$, she cannot deduce out $G_j^{A^\#}$ and $G_j^{B^\#}$ from $R_j^A \oplus G_j^{A^\#}$ and $R_j^B \oplus G_j^{B^\#}$, respectively. Similarly, in Step 3(3), the outside attacker can get other useful classical information including $R_j^A$, $R_j^B$ and the result of $R_j^C \oplus G_j^{C^\#}$. Until now, the outside attacker can extract $G_j^{A^\#}$ and $G_j^{B^\#}$ from $R_j^A \oplus G_j^{A^\#}$ and $R_j^B \oplus G_j^{B^\#}$, respectively, since she has known $R_j^A$ and $R_j^B$. However, the one-way property of the hash function can guarantee that knowing $G_j^{A^\#}$ and $G_j^{B^\#}$ is still helpless to deduce $G_j^A$ and $G_j^B$, respectively. In this way, the outside attacker still has no access to $G_j^A$ and $G_j^B$. On the other hand, the outside attacker cannot get $G_j^{C^\#}$ either since she does not know $R_j^C$. Right now, it can be concluded that an outside eavesdropper cannot get $X$, $Y$ and $Z$ in the three-party QPC protocol.

**Case 2: Participant attack**

Gao F *et al.* [42] pointed out for the first time that the attack from dishonest participants, i.e., the participant attack, is generally more powerful and should be paid more attention to. It has greatly aroused the interest of researchers in the cryptanalysis of quantum cryptography. There are two cases of participant attack in the three-party QPC protocol. The first one is the attack from an insider user, while the second one is the attack from TP.

**(1) Inside user's attack**

Suppose that Alice is a powerful dishonest user who tries her best to get the other users' secret inputs with possible strong means. If Alice tries to intercept the transmitted particles from the TP-Bob channel, the Bob-TP channel, the TP-Charlie channel or the Charlie-TP channel, she will be caught as an outside attacker as analyzed in Case 1. Another way for Alice to get Bob and Charlie's secret inputs is to utilize all the possible classical information in her hands. After the protocol is finished, all the possible classical information Alice has is $R_j^A$, $G_j^{A^\#}$, $R_j^{AB}$, $R_j^B \oplus G_j^{B^\#}$, $R_j^{BC}$, $R_j^B$ and $R_j^{AC}$. As a result, Alice can only deduce out $G_j^{B^\#}$ and $R_j^C \oplus G_j^{C^\#}$ from these classical information, but she still cannot know $G_j^{C^\#}$ since she has no knowledge about $R_j^C$. Moreover, the one-way property of the hash function can make Alice not aware of $G_j^B$ from $G_j^{B^\#}$. Therefore, Alice cannot get $Y$ and $Z$.

Suppose that Bob is a powerful dishonest user who tries his best to get the other users' secret inputs with possible strong means. If Bob tries to intercept the transmitted particles from the TP-Alice channel, the Alice-TP channel, the TP-Charlie channel or the Charlie-TP channel, he will be caught as an outside attacker as analyzed in Case 1. Another way for Bob to get Alice and Charlie's secret inputs is to utilize all the possible classical information in his hands. After the protocol is finished, all the possible classical information Bob has is $R_j^B$, $G_j^{B^\#}$, $R_j^A \oplus G_j^{A^\#}$, $R_j^{AB}$, $R_j^A$, $R_j^{BC}$ and $R_j^{AC}$. As a result, Bob can only deduce out $G_j^{A^\#}$ and $R_j^C \oplus G_j^{C^\#}$ from these classical information, but he still cannot know $G_j^{C^\#}$ since he has no knowledge about $R_j^C$. Moreover, the one-way property of the hash function can make Bob not aware of $G_j^A$ from $G_j^{A^\#}$. Therefore, Bob cannot get $X$ and $Z$.

Suppose that Charlie is a powerful dishonest user who tries her best to get the other users' secret inputs with possible strong means. If Charlie tries to intercept the transmitted particles from the TP-Alice channel, the Alice-TP channel, the TP-Bob channel or the Bob-TP channel, she will be caught as an outside attacker as analyzed in Case 1. Another way for Charlie to get Alice and Bob's secret inputs is to utilize all the possible classical information in her hands. After the protocol is finished, all the possible classical information Charlie has is $R_j^C$, $G_j^{C^\#}$, $R_j^A \oplus G_j^{A^\#}$, $R_j^{AB}$, $R_j^A$, $R_j^B \oplus G_j^{B^\#}$, $R_j^{BC}$, $R_j^B$ and $R_j^{AC}$. As a result, Charlie can deduce out both $G_j^{A^\#}$ and $G_j^{B^\#}$ from these classical information. However, according to the one-way property of the hash function, knowing $G_j^{A^\#}$ and $G_j^{B^\#}$ is still helpless for Charlie to deduce $G_j^A$ and $G_j^B$, respectively. Therefore, Charlie cannot get $X$ and $Y$.

**(2) TP's attack**

TP may try to get Alice, Bob and Charlie's secret inputs with all the possible classical information in her hands. After the protocol is finished, all the possible classical information TP has is $R_j^{T^1}$, $R_j^A \oplus G_j^{A^\#}$, $R_j^{AB}$, $R_j^{AB'}$, $R^{AB}$, $R_j^{T^2}$, $R_j^A$, $R_j^B \oplus G_j^{B^\#}$, $R_j^{BC}$, $R_j^{BC'}$, $R^{BC}$, $R_j^B$, $R_j^{AC}$, $R_j^{AC'}$ and $R^{AC}$. Note that TP needn't launch the Bell-basis measurement attack to get $R_j^A$ and $R_j^B$, as she can get them from the public classical channels. As a result, TP can deduce out all of $G_j^{A^\#}$, $G_j^{B^\#}$ and $G_j^{C^\#}$ from these classical information. However, according to the one-way property of the hash function, knowing $G_j^{A^\#}$, $G_j^{B^\#}$ and $G_j^{C^\#}$ is still helpless for TP to deduce $G_j^A$, $G_j^B$ and $G_j^C$, respectively. Therefore, TP cannot get $X$, $Y$ and $Z$ accurately.

To sum up, in the three-party QPC protocol, TP can know the comparison result of each two users' secret inputs but cannot know the genuine value of each input. Each user cannot know the genuine values of the other two users' secret inputs.

**Case 3: The information leakage problem**

According to formulas (1-3), the relations between $R_j^{AB'}$ and $G_j^{A^\#}$, $G_j^{B^\#}$, the relations between $R_j^{BC'}$ and $G_j^{B^\#}$, $G_j^{C^\#}$, and the relations between $R_j^{AC'}$ and $G_j^{A^\#}$, $G_j^{C^\#}$ can also be depicted as Table 3, respectively. As analyzed in Sec.3.2, the usage of one-way hash function can automatically avoid the information leakage problem pointed out in Sec.3.1.

It can be concluded now that the three-party QPC protocol is highly secure.

## 5 The MQPC protocol based on entanglement swapping of Bell entangled states

There are $K$ users, $P_1$, $P_2$, ..., $P_K$, where $P_i$ has a secret integer $X^i$, $i = 1, 2, ..., K$. The binary representation of $X^i$ in $F_{2^L}$ is $\left(x_{L-1}^i, x_{L-2}^i, ..., x_0^i\right)$. Here, $x_j^i \in \{0,1\}$, $j = 0, 1, ..., L-1$. They want to know whether each two different $X^i$ are equal or not with the help of a semi-honest TP.

They achieve the equality comparison of each two different $X^i$ by implementing the following steps. Same to the above three-party QPC protocol, each transmission of quantum state sequence here is checked with the entanglement correlation between two different particles of a sample Bell state $|\Phi^+\rangle$. For simplicity, we omit the description of eavesdropping check processes in the following.

**Step 1: Preparation.**

(1) Similar to the QPC protocols of Refs.[4-5], $K$ users, $P_1$, $P_2$, ..., $P_K$, share a secret one-way hash

function $H$ in advance. The hash value of $X^i$ is $H(X^i) = X^{i^\#} = (x_{N-1}^{i^\#}, x_{N-2}^{i^\#}, \ldots, x_0^{i^\#})$, $i = 1, 2, \ldots, K$. $P_i$ divides her binary representation of $X^{i^\#}$ into $\lceil N/2 \rceil$ groups $G_1^{P_i^\#}, G_2^{P_i^\#}, \ldots, G_{\lceil N/2 \rceil}^{P_i^\#}$, where each group contains two binary bits. If $N \mod 2 = 1$, one 0 should be added to $G_{\lceil N/2 \rceil}^{P_i^\#}$ by $P_i$.

(2) $P_i$/TP prepares $\lceil N/2 \rceil$ quantum states all in the state of $|\Phi^+\rangle_{P_{i_1} P_{i_2}} / |\Phi^+\rangle_{T_1 T_2}$. Afterward, $P_i$/TP picks out the first particle from each state to form an ordered sequence $S_1^{P_i} / S_1^T$. The remaining second particle from each state automatically forms the other ordered sequence $S_2^{P_i} / S_2^T$.

(3) $P_1$ and TP exchange $S_2^{P_1}$ and $S_2^T$.

(4) For $j = 1, 2, \ldots, \lceil N/2 \rceil$, $P_1$ performs the Bell-basis measurement on each pair in $(S_1^{P_1}, S_2^T)$ and obtains the corresponding measurement result $M_j^{P_1}$. If $M_j^{P_1}$ is $|\Phi^+\rangle / |\Phi^-\rangle / |\Psi^+\rangle / |\Psi^-\rangle$, then $R_j^{P_1} = 00/01/10/11$. Consequently, the corresponding pair in $(S_1^T, S_2^{P_1})$ in TP's hands is collapsed into one of the four Bell states. These $\lceil N/2 \rceil$ collapsed Bell states in TP's hands are denoted by $(S_1^{T^1}, S_2^{T^1})$.

**Step $k$ : The $k-1$ th round comparison $(k = 2, 3, 4, \ldots, K)$.**

(1) $P_k$ and TP exchange $S_2^{P_k}$ and $S_2^{T^{k-1}}$.

(2) For $j = 1, 2, \ldots, \lceil N/2 \rceil$, $P_k$ performs the Bell-basis measurement on each pair in $(S_1^{P_k}, S_2^{T^{k-1}})$ and obtains the corresponding measurement result $M_j^{P_k}$. If $M_j^{P_k}$ is $|\Phi^+\rangle / |\Phi^-\rangle / |\Psi^+\rangle / |\Psi^-\rangle$, then $R_j^{P_k} = 00/01/10/11$. Consequently, the corresponding pair in $(S_1^{T^{k-1}}, S_2^{P_k})$ in TP's hands is collapsed into one of the four Bell states. TP also performs the Bell-basis measurement on each pair in $(S_1^{T^{k-1}}, S_2^{P_k})$ and obtains the corresponding measurement result $M_j^{T^{k-1}}$. If $M_j^{T^{k-1}}$ is $|\Phi^+\rangle / |\Phi^-\rangle / |\Psi^+\rangle / |\Psi^-\rangle$, then $R_j^{T^{k-1}} = (r_j^{T_1^{k-1}} r_j^{T_2^{k-1}}) = 00/01/10/11$.

(3) For $j = 1, 2, \ldots, \lceil N/2 \rceil$, $k$ users cooperate to calculate
$R_j^{P_m P_k} = R_j^{P_1} \oplus R_j^{P_2} \oplus \cdots \oplus R_j^{P_{m-1}} \oplus (R_j^{P_m} \oplus G_j^{P_m^\#}) \oplus R_j^{P_{m+1}} \oplus R_j^{P_{m+2}} \oplus \cdots \oplus R_j^{P_{k-1}} \oplus (R_j^{P_k} \oplus G_j^{P_k^\#}) = (r_j^{P_m P_{k_1}} r_j^{P_m P_{k_2}})$, and send $R_j^{P_m P_k}$ to TP. Here, $m = 1, 2, \ldots, k-1$. Without loss of generality, assume that $P_i$ $(i = 1, 2, \ldots, m-1, m+1, \ldots, k-2, k-1)$ and $P_m$ send $R_j^{P_i}$ and the result of $R_j^{P_m} \oplus G_j^{P_m^\#}$ to $P_k$ for calculating $R_j^{P_m P_k}$, respectively. Then, TP calculates $R_j^{P_m P_k'} = (r_j^{P_m P_{k_1}} \oplus r_j^{T_1^{k-1}}) + (r_j^{P_m P_{k_2}} \oplus r_j^{T_2^{k-1}})$ and $R^{P_m P_k} = \sum_{j=1}^{\lceil N/2 \rceil} R_j^{P_m P_k'}$.

TP sends $R^{P_m P_k}$ to $P_m$ and $P_k$. If $R^{P_m P_k} = 0$, $P_m$ and $P_k$ conclude that $X^m = X^k$; otherwise, they know that $X^m \neq X^k$.

*Correctness.* We continue to demonstrate the output correctness. As for the quality comparison of $X^m$ and $X^k$ ( $m = 1, 2, \ldots, k-1$ and $k = 2, 3, 4, \ldots, K$ ), $k$ users need to calculate $R_j^{P_m P_k} = R_j^{P_1} \oplus R_j^{P_2} \oplus \cdots \oplus R_j^{P_{m-1}} \oplus (R_j^{P_m} \oplus G_j^{P_m^\#}) \oplus R_j^{P_{m+1}} \oplus R_j^{P_{m+2}} \oplus \cdots \oplus R_j^{P_{k-1}} \oplus (R_j^{P_k} \oplus G_j^{P_k^\#}) = (r_j^{P_m P_{k_1}} r_j^{P_m P_{k_2}})$.

Moreover, TP needs to calculate $R_j^{P_m P_k'} = (r_j^{P_m P_{k_1}} \oplus r_j^{T_1^{k-1}}) + (r_j^{P_m P_{k_2}} \oplus r_j^{T_2^{k-1}})$ and $R^{P_m P_k} = \sum_{j=1}^{\lceil N/2 \rceil} R_j^{P_m P_k'}$. According to the entanglement swapping processes of the multi-party QPC protocol, we can obtain

$$\begin{cases} R_{P_{1_1} P_{1_2}} \oplus R_{T_1 T_2} = R_{P_{1_1} T_2} \oplus R_{T_1 P_{1_2}} \\ R_{T_1 P_{1_2}} \oplus R_{P_{2_1} P_{2_2}} = R_{T_1 P_{2_2}} \oplus R_{P_{2_1} P_{1_2}} \\ R_{T_1 P_{2_2}} \oplus R_{P_{3_1} P_{3_2}} = R_{T_1 P_{3_2}} \oplus R_{P_{3_1} P_{2_2}} \\ R_{T_1 P_{3_2}} \oplus R_{P_{4_1} P_{4_2}} = R_{T_1 P_{4_2}} \oplus R_{P_{4_1} P_{3_2}} \\ \quad\quad\quad\quad\quad \vdots \\ R_{T_1 P_{k-1_2}} \oplus R_{P_{k_1} P_{k_2}} = R_{T_1 P_{k_2}} \oplus R_{P_{k_1} P_{k-1_2}} \end{cases}$$

$$\Rightarrow R_{P_{1_1}P_{1_2}} \oplus R_{T_1T_2} = R_{P_{1_1}T_2} \oplus \left( R_{P_{2_1}P_{2_2}} \oplus \left( R_{P_{2_1}P_{1_2}} \oplus \left( R_{P_{3_1}P_{3_2}} \oplus \left( R_{P_{3_1}P_{2_2}} \oplus \cdots \oplus \left( R_{P_{k_1}P_{k_2}} \oplus \left( R_{P_{k_1}P_{k-1_2}} \oplus R_{T_1P_{k_2}} \right)\right)\right)\right)\right)\right)$$

$$\Rightarrow 00 = R_j^{P_1} \oplus R_j^{P_2} \oplus \cdots \oplus R_j^{P_{m-1}} \oplus R_j^{P_m} \oplus R_j^{P_{m+1}} \oplus R_j^{P_{m+2}} \oplus \cdots \oplus R_j^{P_{k-1}} \oplus R_j^{P_k} \oplus R_j^{T^{k-1}}$$

$$\Rightarrow G_j^{P_m^\#} \oplus G_j^{P_k^\#} = R_j^{P_1} \oplus R_j^{P_2} \oplus \cdots \oplus R_j^{P_{m-1}} \oplus \left( R_j^{P_m} \oplus G_j^{P_m^\#} \right) \oplus R_j^{P_{m+1}} \oplus R_j^{P_{m+2}} \oplus \cdots \oplus R_j^{P_{k-1}} \oplus \left( R_j^{P_k} \oplus G_j^{P_k^\#} \right) \oplus R_j^{T^{k-1}}$$

$$= R_j^{P_m P_k} \oplus R_j^{T^{k-1}}$$

$$\Rightarrow R_j^{P_m P_k'} = \begin{cases} 0, & \text{if} \quad G_j^{P_m^\#} = G_j^{P_k^\#}; \\ 1 \text{ or } 2, & \text{if} \quad G_j^{P_m^\#} \neq G_j^{P_k^\#}. \end{cases}$$

$$\Rightarrow R^{P_m P_k} = \sum_{j=1}^{\lceil N/2 \rceil} R_j^{P_m P_k'} = \begin{cases} 0, & \text{if} \quad X^m = X^k; \\ \text{others}, & \text{if} \quad X^m \neq X^k. \end{cases} \qquad (4)$$

Therefore, the quality comparison result of $X^m$ and $X^k$ in the above $K$-party QPC protocol is correct.

**Security.** As far as the security of the MQPC protocol is concerned, we can analyze it in a way similar to that of the three-party QPC protocol. It is easy to find out that the MQPC protocol is also immune to all of the outside attack, the participant attack and the information leakage problem.

***Comparison with previous QPC protocols.*** The comparison of our MQPC protocol with some previous representative QPC protocols, such as Yang Y G *et al.*'s protocol [4], Chen X B *et al.*'s protocol [6], Tseng H Y *et al.*'s protocol[7], Liu W *et al.*'s protocol [8], Yang Y G *et al.*'s protocol [17] and Chang Y J *et al.*'s protocol [35], is described in Table 4. According to Table 4, it is easy to know that each of the protocols in Refs.[4,6-8,17,35] has advantages and disadvantages more or less. For example, our protocol adopts Bell state as quantum resource. As for quantum state used, our protocol takes advantage over the protocols of Refs.[6,35] but is defeated by the protocol of Ref.[17], since the preparation of Bell state is easier than that of GHZ state and is more difficult than that of single photon product state. However, it can be concluded that our protocol exceeds the protocols of Refs.[4,6-8,17] in number of times of protocol execution when they are used to achieve the equality comparison among $K$ users, because in our protocol, arbitrary pair's comparison of equality among $K$ users can be accomplished within one execution.

It should be further emphasized that different quantum methods have been used to achieve the equality comparison in present MQPC protocols [35-37] and our MQPC protocol. Concretely speaking, Chang Y J *et al.*'s protocol [35] uses the entanglement correlation between two different particles of one $n$-particle GHZ class state; both Liu W *et al.*'s protocol [36] and Wang Q L *et al.*'s protocol [37] use quantum fourier transform. However, our protocol uses quantum entanglement swapping.

## 6 Conclusion

In this paper, we first point out the information leakage problem toward TP in the LLCLL-improved-QPC protocol, and then mend this loophole by utilizing the one-way hash function to encrypt the two users' secret inputs. Afterward, the three-party QPC protocol also based on entanglement swapping of Bell entangled state is constructed. Its output correctness and its security against the outside attack, the inside participant attack and the information leakage problem are validated in detail. Finally, the MQPC protocol also based on entanglement swapping of Bell entangled state is designed, where arbitrary pair's comparison of equality among $K$ users can be accomplished within one execution.

**Acknowledgements**

The author would like to thank the anonymous reviewer for his valuable suggestion that helps enhancing the quality of this paper. Funding by the National Natural Science Foundation of China (Grant No.61402407) is also gratefully acknowledged.

**Appendix:**

Table 1.   The relations among different parameters when $G_j^A = 00$

| $G_j^A$ | $G_j^B$ | $M_j^A$ | $M_j^B$ | $R_j^A$ | $R_j^B$ | $R_j \left( r_j^1 r_j^2 \right)$ | $M_j^T$ | $R_j^T \left( r_j^{T_1} r_j^{T_2} \right)$ | $R_j^{'}$ |
|---|---|---|---|---|---|---|---|---|---|
| 00 | 00/01/10/11 | $|\Phi^+\rangle$ | $|\Phi^+\rangle$ | 00 | 00 | 00/01/10/11 | $|\Phi^+\rangle$ | 00 | 0/1/1/2 |
| | | $|\Phi^+\rangle$ | $|\Phi^-\rangle$ | 00 | 01 | 01/00/11/10 | $|\Phi^-\rangle$ | 01 | 0/1/1/2 |
| | | $|\Phi^+\rangle$ | $|\Psi^+\rangle$ | 00 | 10 | 10/11/00/01 | $|\Psi^+\rangle$ | 10 | 0/1/1/2 |
| | | $|\Phi^+\rangle$ | $|\Psi^-\rangle$ | 00 | 11 | 11/10/01/00 | $|\Psi^-\rangle$ | 11 | 0/1/1/2 |
| | | $|\Phi^-\rangle$ | $|\Phi^-\rangle$ | 01 | 01 | 00/01/10/11 | $|\Phi^+\rangle$ | 00 | 0/1/1/2 |
| | | $|\Phi^-\rangle$ | $|\Phi^+\rangle$ | 01 | 00 | 01/00/11/10 | $|\Phi^-\rangle$ | 01 | 0/1/1/2 |
| | | $|\Phi^-\rangle$ | $|\Psi^-\rangle$ | 01 | 11 | 10/11/00/01 | $|\Psi^+\rangle$ | 10 | 0/1/1/2 |
| | | $|\Phi^-\rangle$ | $|\Psi^+\rangle$ | 01 | 10 | 11/10/01/00 | $|\Psi^-\rangle$ | 11 | 0/1/1/2 |
| | | $|\Psi^+\rangle$ | $|\Psi^+\rangle$ | 10 | 10 | 00/01/10/11 | $|\Phi^+\rangle$ | 00 | 0/1/1/2 |
| | | $|\Psi^+\rangle$ | $|\Psi^-\rangle$ | 10 | 11 | 01/00/11/10 | $|\Phi^-\rangle$ | 01 | 0/1/1/2 |
| | | $|\Psi^+\rangle$ | $|\Phi^+\rangle$ | 10 | 00 | 10/11/00/01 | $|\Psi^+\rangle$ | 10 | 0/1/1/2 |

| | | | | | | | |
|---|---|---|---|---|---|---|---|
| $\lvert\Psi^+\rangle$ | $\lvert\Phi^-\rangle$ | 10 | 01 | 11/10/01/00 | $\lvert\Psi^-\rangle$ | 11 | 0/1/1/2 |
| $\lvert\Psi^-\rangle$ | $\lvert\Phi^+\rangle$ | 11 | 00 | 11/10/01/00 | $\lvert\Psi^-\rangle$ | 11 | 0/1/1/2 |
| $\lvert\Psi^-\rangle$ | $\lvert\Psi^+\rangle$ | 11 | 10 | 01/00/11/10 | $\lvert\Phi^-\rangle$ | 01 | 0/1/1/2 |
| $\lvert\Psi^-\rangle$ | $\lvert\Phi^-\rangle$ | 11 | 01 | 10/11/00/01 | $\lvert\Psi^+\rangle$ | 10 | 0/1/1/2 |
| $\lvert\Psi^-\rangle$ | $\lvert\Psi^-\rangle$ | 11 | 11 | 00/01/10/11 | $\lvert\Phi^+\rangle$ | 00 | 0/1/1/2 |

Table 2.   The relations between $R'_j$ and $G^A_j$, $G^B_j$

| $R'_j$ | $G^A_j$ | $G^B_j$ |
|---|---|---|
| 0 | 00/01/10/11 | 00/01/10/11 |
| 1 | 00/01/10/11 | 01/00/11/10 |
| | | 10/11/00/01 |
| 2 | 00/01/10/11 | 11/10/01/00 |

Table 3.   The relations between $R'_j$ and $G^{A^\#}_j$, $G^{B^\#}_j$

| $R'_j$ | $G^{A^\#}_j$ | $G^{B^\#}_j$ |
|---|---|---|
| 0 | 00/01/10/11 | 00/01/10/11 |
| 1 | 00/01/10/11 | 01/00/11/10 |
| | | 10/11/00/01 |
| 2 | 00/01/10/11 | 11/10/01/00 |

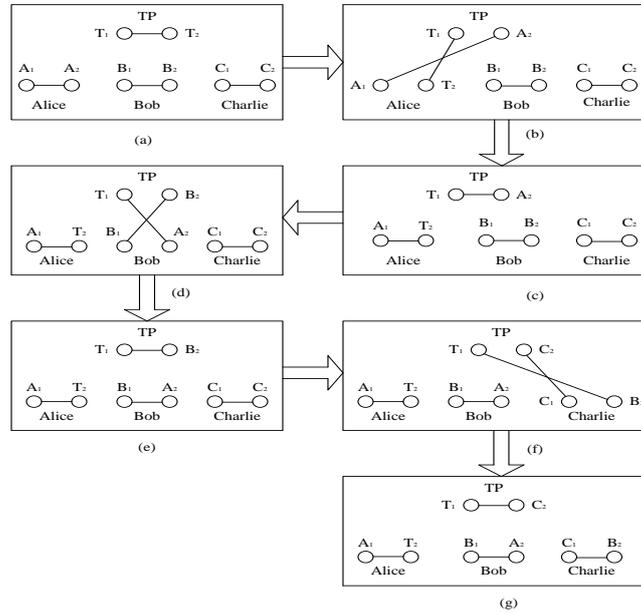

Fig.1.   The entanglement swapping of Bell states among the four participants. (a)Alice/Bob/Charlie/TP prepares quantum states in the state of $\lvert\Phi^+\rangle_{A_1A_2}/\lvert\Phi^+\rangle_{B_1B_2}/\lvert\Phi^+\rangle_{C_1C_2}/\lvert\Phi^+\rangle_{T_1T_2}$. (b) Alice and TP exchange the second particles $A_2$ and $T_2$ of the Bell states in their respective hands. (c) Particles $T_1$ and $A_2$ in TP's hands become entangled together after Alice performs the Bell-basis measurement on particles $A_1$ and $T_2$. (d) TP and Bob exchange particles $A_2$ and $B_2$. (e) Particles $T_1$ and $B_2$ in TP's hands become entangled together after Bob performs the Bell-basis measurement on particles $B_1$ and $A_2$. (f) TP and Charlie exchange particles $B_2$ and $C_2$. (g) Particles $T_1$ and $C_2$ in TP's hands become entangled together after Charlie performs the Bell-basis measurement on particles $C_1$ and $B_2$.

Table 4.   The comparison of our MQPC protocol with previous QPC protocols

| Yang Y G et al.'s protocol [4] | Chen X B et al.'s protocol [6] | Tseng H Y et al.'s protocol [7] | Liu W et al.'s protocol [8] | Yang Y G et al.'s protocol [17] | Chang Y J et al.'s protocol [35] | Our protocol |
|---|---|---|---|---|---|---|

| Quantum state | Bell state | Triple GHZ state | Bell state | Bell state | Single-photon product state | n-particle GHZ class state | Bell state |
|---|---|---|---|---|---|---|---|
| Ye T Y: Multi-party Quantum Private Comparison Protocol Based on Entanglement Swapping of Bell Entangled States | | | | | | | |
| Quantum measurement for TP | Bell-basis measurement | Single-photon measurement | No | Bell-basis measurement | No | No | Bell-basis measurement |
| Quantum measurement for users | No | Single-photon measurement | Single-photon measurement | Bell-basis measurement | Single-photon measurement | Single-photon measurement | Bell-basis measurement |
| Unitary operation for TP | No | Yes | No | No | No | No | No |
| Unitary operation for users | Yes | No | No | No | No | No | No |
| Quantum memory for TP | No | Yes | No | Yes | No | No | Yes |
| Number of times of protocol execution | $(K-1) \sim K(K-1)/2$ | $(K-1) \sim K(K-1)/2$ | $(K-1) \sim K(K-1)/2$ | $(K-1) \sim K(K-1)/2$ | $(K-1) \sim K(K-1)/2$ | 1 | 1 |